# Query strategy for sequential ontology debugging[☆]

Kostyantyn Shchekotykhin[a,*], Gerhard Friedrich[a], Philipp Fleiss[a], Patrick Rodler[a]

[a]*Alpen-Adria Universität, Universitätsstrasse 65-67, 9020 Klagenfurt, Austria*

**Abstract**

Debugging of ontologies is an important prerequisite for their wide-spread application, especially in areas that rely upon everyday users to create and maintain knowledge bases, as in the case of the Semantic Web. Recent approaches use diagnosis methods to identify causes of inconsistent or incoherent ontologies. However, in most debugging scenarios these methods return many alternative diagnoses, thus placing the burden of fault localization on the user. This paper demonstrates how the target diagnosis can be identified by performing a sequence of observations, that is, by querying an oracle about entailments of the target ontology. We exploit a-priori probabilities of typical user errors to formulate information-theoretic concepts for query selection. Our evaluation showed that the proposed method significantly reduces the number of required queries compared to myopic strategies. We experimented with different probability distributions of user errors and different qualities of the a-priori probabilities. Our measurements showed the advantageousness of information-theoretic approach to query selection even in cases where only a rough estimate of the priors is available.

*Keywords:* Ontology Debugging, Query Selection, Model-based Diagnosis, Description Logic

## 1. Introduction

Acquisition and maintenance of knowledge bases is an important prerequisite for a successful application of semantic systems in areas such as the Semantic Web. At the current state of the art ontology extraction methods do not allow a complete and error free automatic acquisition of ontologies. Thus users of semantic systems are required to formulate and correct logical descriptions on their own. In most of the cases these users are domain experts who have little or no experience in expressing their knowledge in representation languages like OWL [1]. Studies in cognitive psychology, e.g. [2, 3], discovered that humans make systematic errors while formulating or interpreting logical descriptions. Results presented in [4, 5] confirmed these observations regarding ontology development. Therefore it is essential to create methods that can identify and correct erroneous ontological definitions.

Ontology debugging methods [6, 7, 8, 9] simplify the development of ontologies. Usually the main requirement for the debugging process is to obtain a consistent and, optionally, coherent ontology. These basic requirements can be extended by additional ones, such as test cases [8], which must be fulfilled by the target ontology $O_t$. Given the requirements (e.g. formulated by a user) an ontology debugger identifies a set of alternative diagnoses, where each diagnosis corresponds to a set of possibly faulty axioms. In particular, a *diagnosis* $\mathcal{D}$ is a subset of an ontology $O$ such that removal of the diagnosis from the ontology (i.e. $O \setminus \mathcal{D}$) will allow the formulation of the target ontology $O_t$ that fulfills all the requirements. We call the removal of a diagnosis from the ontology a *trivial application* of a diagnosis. Moreover, in practical applications it might be inefficient to consider all possible diagnoses. Therefore, modern ontology debugging approaches focus on the computation of minimal diagnoses, i.e. such diagnoses $\mathcal{D}_i$ that no $\mathcal{D}'_i \subset \mathcal{D}_i$ is a diagnosis. A user has to change at least all of the axioms of a minimal diagnosis in order to formulate the intended target ontology.

However, the diagnosis methods can return many alternative minimal diagnoses for a given set of test cases

[*]Corresponding author at: Alpen-Adria Universität, Universitätsstrasse 65-67, 9020 Klagenfurt, Austria. Tel: +43 463 2700 3768, Fax:++43 463 2700 993768

*Email addresses:* kostya@ifit.uni-klu.ac.at (Kostyantyn Shchekotykhin), gerhard@ifit.uni-klu.ac.at (Gerhard Friedrich), pfleiss@ifit.uni-klu.ac.at (Philipp Fleiss), prodler@ifit.uni-klu.ac.at (Patrick Rodler)

and requirements. A sample study of real-world incoherent ontologies, which were used in [7], shows that there may exist hundreds or even thousands of minimal diagnoses. In the case of the Transportation ontology the diagnosis method was able to identify 1782 minimal diagnoses [1]. In such situations some simple visualization of all alternative modifications of the ontology is ineffective. The goal of sequential debugging is to identify the set of axioms $\mathcal{D}_t$ of an ontology which have to be changed or removed in order to formulate the target ontology $O_t$. The set of axioms $\mathcal{D}_t$ is called the target diagnosis. Consequently, the target ontology corresponds to the ontology resulting from a removal of the target diagnosis from the original ontology and an extension by some additional axioms $EX$, i.e. $O_t = (O \setminus \mathcal{D}_t) \cup EX$.

A possible solution of the problem would be to introduce an order on the set of diagnoses by means of some preference criteria. For instance, Kalyanpur et al. [10] suggest measures to rank the axioms of a diagnosis depending on their structure, occurrence in test cases, etc. Only the top ranking diagnoses are then presented to the user. Of course this set of diagnoses will contain the target diagnosis only in the case when a faulty ontology, the given requirements and test cases, provide sufficient data to the appropriate heuristic. Therefore, in most debugging sessions a user has to input additional information (e.g. tests in form of required implications of facts or axioms) to identify the target diagnosis. However, it is hard to guess, which information is required. That is, a user does not know a priori which and how many tests should be provided to the debugger, such that it will return the target diagnosis.

In this paper we present an approach for the acquisition of additional information by generating a sequence of queries, which should be answered by some oracle such as a user, an information extraction system, etc. Each answer to a query is used by our method to reduce the set of diagnoses until, finally, the target diagnosis is identified. In order to construct queries we exploit the property that different ontologies resulting from trivial applications of different diagnoses entail unequal sets of axioms. Consequently, we can differentiate between diagnoses by asking the oracle if the target ontology should imply a logical sentence or not. These implied logical sentences can be generated by classification and realization services provided in description logic reasoning systems [11, 12, 13]. In particular, the classification process computes a subsumption hierarchy (sometimes also called "inheritance hierarchy" of parents and children) for each concept name mentioned in a TBox. For each individual mentioned in an ABox, the realization computes the atomic classes (or concept names) of which the individual is an instance [11].

In order to generate the most informative query we exploit the fact that some diagnoses are more likely than others because of typical user errors [4, 5]. User's beliefs for an error to occur in some part of a knowledge base, represented as probabilities, can be used to estimate the change in entropy of the set of diagnoses if a particular query is answered. We select those queries which minimize the expected entropy, i.e. maximize the information gain. An oracle should answer these queries until a diagnosis is identified whose probability is significantly higher than those of all other diagnoses. This diagnosis is the most likely to be the target one.

We compare our entropy-based method with a greedy approach that selects those queries which try to cut the number of diagnoses in half. The evaluation is performed using generated examples as well as real-world ontologies presented in Table 8. In the first case we alter a consistent and coherent ontology with additional axioms to generate such conflicts that result in a predefined number of diagnoses of required length. A faulty ontology is then analyzed by the debugging algorithm using entropy, greedy and "random" strategies, where the latter selects queries to be asked completely randomly. Evaluation results show that on average the suggested entropy-based approach is almost 50% better than the greedy one. In the second evaluation scenario we analyzed the performance of entropy-based and greedy strategies on real-world ontologies given different input settings. In particular, we simulated different strategies for a user to assign prior probabilities as well as the quality of these probabilities that might occur in practice. The obtained results show that the entropy method outperformed the greedy heuristic in most of the cases. In some situations the entropy-based approach achieved twice as good average performance compared to the greedy one. Moreover, the evaluation on the real-world ontologies showed that the entropy-based query selection is robust to the actual values of prior fault probabilities as well as differences between them. It is only important whether the specified priors favor the target diagnosis or not.

The remainder of the paper is organized as follows: Section 2 presents two introductory examples as well as the basic concepts. The details of the entropy-based query selection method are given in Section 3. Section 4 describes the implementation of the approach and is followed by evaluation results in Section 5. The paper concludes with an overview on related work.

[1]Subsequently, we will give a detailed characterization of these ontologies.

## 2. Motivating examples and basic concepts

First, we present the fundamental concepts regarding the diagnosis of ontologies and eventually show how queries and answers can be generated and employed to differentiate between sets of diagnoses.

### 2.1. Diagnosis of ontologies

**Example 1.** *Consider a simple ontology $O$ with the terminology $\mathcal{T}$:*

$$ax_1 : A \sqsubseteq B \quad ax_2 : B \sqsubseteq C$$
$$ax_3 : C \sqsubseteq D \quad ax_4 : D \sqsubseteq R$$

*and assertions* $\mathcal{A} : \{A(w), \neg R(w), A(v)\}$.

Let the user explicitly define that the three assertional axioms should be considered as correct, i.e. these axioms are added to a background theory $\mathcal{B}$. The introduction of a background theory keeps the diagnosis method focused on the possibly faulty axioms.

Assume that the user requires the ontology $O$ to be consistent, whereas $O$ is inconsistent. The only irreducible set of axioms (minimal conflict set) that preserves the inconsistency is $CS : \{\langle ax_1, ax_2, ax_3, ax_4\rangle\}$. That is one has to modify or remove the axioms of at least one of the following diagnoses

$$\mathcal{D}_1 : [ax_1] \quad \mathcal{D}_2 : [ax_2] \quad \mathcal{D}_3 : [ax_3] \quad \mathcal{D}_4 : [ax_4]$$

to restore the consistency of the ontology. However, it is unclear, which diagnosis from the set $\mathbf{D} : \{\mathcal{D}_1, \dots, \mathcal{D}_4\}$ corresponds to the target one.

The target diagnosis can be identified by the debugger given a set of axioms $P$ that *must* be entailed by the target ontology and a set of axioms $N$ that *must not*:

1. $O_t \models p \; \forall p \in P$
2. $O_t \not\models n \; \forall n \in N$

For instance, if the user provides the information that $O_t \models B(w)$ and $O_t \not\models C(w)$ then the debugger will return only one diagnosis in our example, namely $\mathcal{D}_2$. Application of this diagnosis results in a satisfiable ontology $O_2 = O \setminus \mathcal{D}_2$ that entails $B(w)$ because of $ax_1$ and the assertion $A(w)$. In addition, $O_2$ does not entail $C(w)$ since $O_2 \sqcap \neg C(w)$ is satisfiable and, moreover, $\neg R(w) \sqcap ax_4 \sqcap ax_3 \models \neg C(w)$. All other ontologies $O_i = (O \setminus \mathcal{D}_i)$ obtained by the application of the diagnoses $\mathcal{D}_1, \mathcal{D}_3$ and $\mathcal{D}_4$ do not fulfill the given requirements, since $O_1 \cup B(w)$ is unsatisfiable and therefore any satisfiable extension of $O_1$ cannot entail $B(w)$ and both $O_3$ and $O_4$ entail $C(w)$. Therefore, $O_2$ corresponds to the target diagnosis $O_t$.

Note that the approach presented in this paper can also be used with knowledge representation languages without negation like *OWL 2 EL* if an underlying reasoner supports both consistency and entailment checking.

**Definition 1.** *Given a diagnosis problem instance $\langle O, \mathcal{B}, P, N\rangle$ where $O$ is an ontology, $\mathcal{B}$ a background theory, $P$ a set of logical sentences which must be implied by the target ontology $O_t$, and $N$ a set of logical sentences which must* not *be implied by $O_t$.*

*A diagnosis is a set of axioms $\mathcal{D} \subseteq O$ iff the set of axioms $O \setminus \mathcal{D}$ can be extended by a logical description $EX$ such that:*

1. *$(O \setminus \mathcal{D}) \cup \mathcal{B} \cup EX$ is consistent (and coherent if required by a user)*
2. *$(O \setminus \mathcal{D}) \cup \mathcal{B} \cup EX \models p$ for all $p \in P$*
3. *$(O \setminus \mathcal{D}) \cup \mathcal{B} \cup EX \not\models n$ for all $n \in N$*

Following the standard definition of diagnosis [14, 15], it is assumed that each axiom $ax_j \in \mathcal{D}_i$ is faulty, whereas each axiom $ax_k \in O \setminus \mathcal{D}_i$ is correct.

If $\mathcal{D}_t$ is the set of axioms of $O$ to be changed (i.e. $\mathcal{D}_t$ is the target diagnosis) then the target ontology $O_t$ is $(O \setminus \mathcal{D}_t) \cup \mathcal{B} \cup EX$ for some $EX$ defined by the user.

**Definition 2.** *A diagnosis $\mathcal{D}$ for a diagnosis problem instance $\langle O, \mathcal{B}, P, N\rangle$ is a* minimal diagnosis *iff there is no proper subset of the faulty axioms $\mathcal{D}' \subset \mathcal{D}$ such that $\mathcal{D}'$ is a diagnosis.*

**Definition 3.** *A diagnosis $\mathcal{D}$ for a diagnosis problem instance $\langle O, \mathcal{B}, P, N\rangle$ is a* minimum cardinality diagnosis *iff there is no diagnosis $\mathcal{D}' \subset \mathcal{D}$ such that $|\mathcal{D}'| < |\mathcal{D}|$.*

The extension $EX$ plays an important role in the repair process of an ontology. A diagnosis suggests only some set of axioms, which have to be removed from an ontology by the user, but it does not make any suggestion on axioms that have to be added to the ontology. For instance, given our example ontology $O$, the user requires that the target ontology *must not* entail $B(w)$ but has to entail $B(v)$, that is $N = \{B(w)\}$ and $P = \{B(v)\}$. Because, the example ontology is inconsistent some sentences must be changed. The consistent ontology $O_1 = O \setminus \mathcal{D}_1$, neither entails $B(v)$ nor $B(w)$ (in particular $O_1 \models \neg B(w)$). Consequently, $O_1$ has to be extended with some set $EX$ of logical sentences in order to entail $B(v)$. This set of logical sentences can be simply approximated with $EX = \{B(v)\}$. $O_1 \cup EX$ is satisfiable, entails $B(v)$ but does not entail $B(w)$.

All other ontologies $O_i = O \setminus \mathcal{D}_i, \; i = 2, 3, 4$ are consistent but entail both $B(w)$ and $B(v)$ and must be rejected because of the monotonic semantic of description logic. That is, there is no such extension $EX$ that $(O_i \cup EX) \not\models B(w)$. Therefore, the diagnosis $\mathcal{D}_1$ is the minimum cardinality diagnosis which allows the formulation of the target ontology with changing a minimal number of axioms.

The following proposition characterizes diagnoses without the true extension $EX$ employed to formulate the target ontology. The idea is to use the sentences which must be entailed by the target ontology to approximate $EX$ as it is shown above.

**Corollary 1.** *Given a diagnosis problem $\langle O, \mathcal{B}, P, N\rangle$, a set of axioms $\mathcal{D} \subseteq O$ is a diagnosis iff*

$$(O \setminus \mathcal{D}) \cup \mathcal{B} \cup \{\bigwedge_{p \in P} p\}$$

*is satisfiable (coherent) and*

$$\forall n \in N \ : \ (O \setminus \mathcal{D}) \cup \mathcal{B} \cup \{\bigwedge_{p \in P} p\} \not\models n$$

In the following we assume that a diagnosis always exists. A diagnosis exists iff the background theory together with the axioms in $P$ are consistent (coherent) and no axiom in $N$ is entailed, i.e.

**Proposition 1.** *A diagnosis $\mathcal{D}$ for a diagnosis problem $\langle O, \mathcal{B}, P, N\rangle$ exists iff*

$$\mathcal{B} \cup \{\bigwedge_{p \in P} p\}$$

*is consistent (coherent) and*

$$\forall n \in N \ : \ \mathcal{B} \cup \{\bigwedge_{p \in P} p\} \not\models n$$

For the computation of diagnoses *conflict sets* are usually employed to constrain the search space. A conflict set is the part of the ontology that preserves the inconsistency/incoherency.

**Definition 4.** *Given a diagnosis problem instance $\langle O, \mathcal{B}, P, N\rangle$, a set of axioms $CS \subseteq O$ is a conflict set iff $CS \cup \mathcal{B} \cup \{\bigwedge_{p \in P} p\}$ is inconsistent (incoherent) or there is an $n \in N$ s.t. $CS \cup \mathcal{B} \cup \{\bigwedge_{p \in P} p\} \models n$.*

**Definition 5.** *A conflict set $CS$ for an instance $\langle O, \mathcal{B}, P, N\rangle$ is minimal iff there is no proper subset $CS' \subset CS$ such that $CS'$ is a conflict.*

A set of minimal conflict sets can be used to compute the set of minimal diagnoses as it is shown in [14]. The idea is that each diagnosis should include at least one element of each minimal conflict set.

| Ontology | Entailments |
|---|---|
| $O_1$ | $\emptyset$ |
| $O_2$ | $\{B(w)\}$ |
| $O_3$ | $\{B(w), C(w)\}$ |
| $O_4$ | $\{B(w), C(w), D(w)\}$ |

Table 1: Entailments of ontologies $O_i = (O \setminus \mathcal{D}_i)$ , $i = 1 \ldots 4$ in Example 1 returned by realization.

**Proposition 2.** *$\mathcal{D}$ is a diagnosis for the diagnosis problem instance $\langle O, \mathcal{B}, P, N\rangle$ iff $\mathcal{D}$ is a minimal hitting set for the set of all minimal conflict sets of the instance.*

Most of the modern ontology diagnosis methods [6, 7, 8, 9] are implemented according to Proposition 2 and differ in details, e.g. how and when (minimal) conflict sets are computed, the order in which hitting sets are generated, etc.

### 2.2. Differentiating between diagnoses

The diagnosis method usually generates a set of diagnoses for a given diagnosis problem instance. Thus, in Example 1 an ontology debugger returns four minimal diagnoses $\{\mathcal{D}_1 \ldots \mathcal{D}_4\}$. As it is shown in the previous section, additional information, i.e. sets of logical sentences $P$ and $N$, can be used by the debugger to reduce the set of diagnoses. However, in the general case the user does not know which sets $P, N$ of logical sentences should be provided to the debugger s.t. the target diagnosis is identified. Therefore, the debugger should be able to identify sets of logical sentences on its own and only ask the user or some other oracle, whether these sentences *must* or *must not* be entailed by the target ontology. To generate these sentences the debugger can apply each of the diagnoses $\mathbf{D} = \{\mathcal{D}_1 \ldots \mathcal{D}_n\}$ and obtain a set of ontologies $O_i = O \setminus \mathcal{D}_i$ that fulfill the user requirements. For every ontology $O_i$ a description logic reasoner can generate a set of entailments such as entailed subsumptions provided by the classification service and sets of class assertions provided by the realization. In fact, the intention of the classification is that a model for a specific application domain can be verified by exploiting the subsumption hierarchy [16]. These entailments can be used to discriminate between the diagnoses, as different ontologies are likely to entail different sets of sentences. In the following we consider only two types of entailments that can be computed by a description logic reasoner, namely subsumptions and class assertions. In general, the approach presented in this paper is not limited to these types and can use all possible entailment types supported by a reasoner.

For instance, in Example 1 for each ontology $O_i = (O \setminus \mathcal{D}_i)$, $i = 1 \dots 4$ the realization service of a reasoner returns the set of class assertions presented in Table 1. Without any additional information the debugger cannot decide which of these sentences must be entailed by the target ontology. To get this information the diagnosis method should be able to access some oracle that can answer whether the target ontology entails some set of sentences or not. E.g. the debugger asks an oracle if $D(w)$ is entailed by the target ontology ($O_t \models D(w)$). If the answer is *yes*, then $D(w)$ is added to $P$ and $\mathcal{D}_4$ is considered as the target diagnosis. All other diagnoses are rejected because $(O \setminus \mathcal{D}_i) \cup \mathcal{B} \cup \{D(w)\}$ for $i = 1, 2, 3$ is inconsistent. If the answer is *no*, then $D(w)$ is added to $N$ and $\mathcal{D}_4$ is rejected as $(O \setminus \mathcal{D}_4) \cup \mathcal{B} \models D(w)$ and we have to ask the oracle another question.

**Property 1.** *Given a diagnosis problem $\langle O, \mathcal{B}, P, N \rangle$, a set of diagnoses $\mathbf{D}$, and a set of logical sentences $Q$ representing the query $O_t \models Q$ :*

*If the oracle gives the answer* yes *then every diagnosis $\mathcal{D}_i \in \mathbf{D}$ is a diagnosis for $P \cup Q$ iff both conditions hold:*

$$(O \setminus \mathcal{D}_i) \cup \mathcal{B} \cup \{\bigwedge_{p \in P} p\} \cup Q \text{ is consistent (coherent)}$$

$$\forall n \in N \ : \ (O \setminus \mathcal{D}_i) \cup \mathcal{B} \cup \{\bigwedge_{p \in P} p\} \cup Q \not\models n$$

*If the oracle gives the answer* no *then every diagnosis $\mathcal{D}_i \in \mathbf{D}$ is a diagnosis for $N \cup Q$ iff both conditions hold:*

$$(O \setminus \mathcal{D}_i) \cup \mathcal{B} \cup \{\bigwedge_{p \in P} p\} \text{ is consistent (coherent)}$$

$$\forall n \in (N \cup Q) \ : \ (O \setminus \mathcal{D}_i) \cup \mathcal{B} \cup \{\bigwedge_{p \in P} p\} \not\models n$$

In particular, a query partitions the set of diagnoses $\mathbf{D}$ into three mutual disjoint subsets.

**Definition 6.** *For a query $Q$ each diagnosis $\mathcal{D}_i \in \mathbf{D}$ of a diagnosis problem instance $\langle O, \mathcal{B}, P, N \rangle$ can be assigned to one of the three sets $\mathbf{D^P}$, $\mathbf{D^N}$ or $\mathbf{D^\emptyset}$ where*

- *$\mathcal{D}_i \in \mathbf{D^P}$ if it holds that*

$$(O \setminus \mathcal{D}_i) \cup \mathcal{B} \cup \{\bigwedge_{p \in P} p\} \models Q$$

- *$\mathcal{D}_i \in \mathbf{D^N}$ if it holds that*

$$(O \setminus \mathcal{D}_i) \cup \mathcal{B} \cup \{\bigwedge_{p \in P} p\} \cup Q$$

  *is inconsistent (incoherent).*

- *$\mathcal{D}_i \in \mathbf{D^\emptyset}$ if $\mathcal{D}_i \notin \left(\mathbf{D^P} \cup \mathbf{D^N}\right)$*

Given a diagnosis problem instance we say that the diagnoses in $\mathbf{D^P}$ predict a positive answer (*yes*) as a result of the query $Q$, diagnoses in $\mathbf{D^N}$ predict a negative answer (*no*), and diagnoses in $\mathbf{D^\emptyset}$ do not make any predictions.

**Property 2.** *Given a diagnosis problem instance $\langle O, \mathcal{B}, P, N \rangle$, a set of diagnoses $\mathbf{D}$, and a query $Q$:*

*If the oracle gives the answer* yes *then the set of rejected diagnoses is $\mathbf{D^N}$ and the set of remaining diagnoses is $\mathbf{D^P} \cup \mathbf{D^\emptyset}$.*

*If the oracle gives the answer* no *then the set of rejected diagnoses is $\mathbf{D^P}$ and the set of remaining diagnoses is $\mathbf{D^N} \cup \mathbf{D^\emptyset}$.*

Consequently, given a query $Q$ either $\mathbf{D^P}$ or $\mathbf{D^N}$ are eliminated but $\mathbf{D^\emptyset}$ always remains after the query is answered. For generating queries we have to investigate for which subsets $\mathbf{D^P}, \mathbf{D^N} \subseteq \mathbf{D}$ a query exists that can differentiate between these sets. A straight forward approach for query generation is to investigate all possible subsets of $\mathbf{D}$. This is feasible if we limit the number $n$ of minimal diagnoses to be considered during query generation and selection. E.g. for $n = 9$ in the worst case the algorithm has to verify 512 possible partitions.

Given a set of diagnoses $\mathbf{D}$ for the ontology $O$, a set $P$ of sentences that must be entailed by the target ontology $O_t$ and a set of background axioms $\mathcal{B}$, the set of partitions $\mathbf{PR}$ for which a query exists can be computed as follows:

1. Generate the power set $\mathcal{P}(\mathbf{D})$, $\mathbf{PR} \leftarrow \emptyset$
2. Assign to the set $\mathbf{D^P_i}$ an element of $\mathcal{P}(\mathbf{D})$ and generate a set of common entailments $E_i$ of all ontologies $O \setminus \mathcal{D}_j$, where $\mathcal{D}_j \in \mathbf{D^P_i}$
3. If $E_i = \emptyset$ then reject the current element, remove it from $\mathcal{P}(\mathbf{D}) \leftarrow \mathcal{P}(\mathbf{D}) \setminus \mathbf{D^P_i}$ and goto Step 2. Otherwise set $Q_i \leftarrow E_i$.
4. Use Definition 6 and the query $Q_i$ to classify the diagnoses $\mathcal{D}_k \in \mathbf{D} \setminus \mathbf{D^P_i}$ into the sets $\mathbf{D^P_i}$, $\mathbf{D^N_i}$ and $\mathbf{D^\emptyset_i}$. The generated partition is added to the set of partitions $\mathbf{PR} \leftarrow \mathbf{PR} \cup \{\left\langle Q_i, \mathbf{D^P_i}, \mathbf{D^N_i}, \mathbf{D^\emptyset_i} \right\rangle\}$ and set $\mathcal{P}(\mathbf{D}) \leftarrow \mathcal{P}(\mathbf{D}) \setminus \mathbf{D^P_i}$. If $\mathcal{P}(\mathbf{D}) \neq \emptyset$ then goto Step 2.

In Example 1 the set of diagnoses $\mathbf{D}$ of the ontology $O$ contains 4 elements. Therefore, the power set $\mathcal{P}(\mathbf{D})$ includes 16 elements $\{\{\mathcal{D}_1\}, \{\mathcal{D}_2\}, \dots, \{\mathcal{D}_1, \mathcal{D}_2, \mathcal{D}_3, \mathcal{D}_4\}\}$. However, we can omit the element corresponding to $\emptyset$ as it does not contains any diagnoses to be evaluated. Moreover, assume that $P$ and $N$ are empty. On each iteration an element of $\mathcal{P}(\mathbf{D})$ is assigned to the set $\mathbf{D^P_i}$. For instance, the algorithm assigned $\mathbf{D_1^P} = \{\mathcal{D}_1, \mathcal{D}_2\}$. In this case the set of common entailments is empty as $O \setminus \mathcal{D}_1$ has no entailed instances (in addition to the given class assertions, see Table 1). Therefore, the set $\{\mathcal{D}_1, \mathcal{D}_2\}$ is rejected and removed from $\mathcal{P}(\mathbf{D})$. Assume that on the next iteration the algorithm selected $\mathbf{D_2^P} = \{\mathcal{D}_2, \mathcal{D}_3\}$. In this case the set of common entailments $E_2 = \{B(w)\}$ is not empty and so $Q_2 = \{B(w)\}$. The remaining diagnoses $\mathcal{D}_1$ and $\mathcal{D}_4$ are classified according to Definition 6. That is, the algorithm selects the first diagnosis $\mathcal{D}_1$ and verifies whether $(O \setminus \mathcal{D}_1) \models \{B(w)\}$. Given the negative answer of the reasoner, the algorithm checks if $(O \setminus \mathcal{D}_1) \cup \{B(w)\}$ is inconsistent. Since the condition is satisfied the diagnosis $\mathcal{D}_1$ is added to the set $\mathbf{D_2^N}$. The second diagnosis $\mathcal{D}_4$ is added to the set $\mathbf{D_2^P}$ as it satisfies the first requirement $(O \setminus \mathcal{D}_4) \models \{B(w)\}$. The resulting partition $\langle\{B(w)\}, \{\mathcal{D}_2, \mathcal{D}_3, \mathcal{D}_4\}, \{\mathcal{D}_1\}, \emptyset\rangle$ is added to the set **PR**.

However, a query need not include all of the entailed sentences. If a query $Q$ partitions the set of diagnoses into $\mathbf{D^P}$, $\mathbf{D^N}$ and $\mathbf{D^\emptyset}$ and there exists an (irreducible) subset $Q' \subset Q$ which preserves the partition then it is sufficient to query $Q'$. In our example, $Q_2 : \{B(w), C(w)\}$ can be reduced to its subset $Q'_2 : \{C(w)\}$. If there are multiple irreducible subsets that preserve the partition then we select one of them.

All queries and corresponding partitions generated in Example 1 are presented in Table 2. Given these queries the debugger has to decide which one should be asked first in order to minimize the number of queries to be answered. A popular query selection heuristic (called "Split-in-half") prefers those queries, which allow to remove a half of the diagnoses from the set **D**, regardless of the answer of an oracle.

Using the data presented in Table 2, the "Split-in-half" heuristic determines that asking the oracle if $O_t \models \{C(w)\}$ is the best query (i.e. the reduced query $Q_2$), as two diagnoses from the set **D** are removed regardless of the answer. Let us assume that $\mathcal{D}_1$ is the target diagnosis, then an oracle will answer *no* to our question (i.e. $O_t \not\models \{C(w)\}$). Based on this feedback, the diagnoses $\mathcal{D}_3$ and $\mathcal{D}_4$ are removed according to Property 2. Given the updated set of diagnoses **D** and $P = \{C(w)\}$ the partitioning algorithm returns the only partition $\langle\{B(w)\}, \{\mathcal{D}_2\}, \{\mathcal{D}_1\}, \emptyset\rangle$. Therefore we ask the query $\{B(w)\}$, which is also answered with *no* by the oracle. Consequently, we identified $\mathcal{D}_1$ as the only remaining minimal diagnosis.

In general, if $n$ is the number of diagnoses and we can split the set of diagnoses in half by each query, then the minimum number of queries is $log_2 n$. However, if the probabilities of diagnoses are known we can reduce this number of queries by using two effects:

1. We can exploit diagnoses probabilities to asses the probabilities of answers and the expected value of information contained in the set of diagnoses after an answer is given.
2. Even if there are multiple diagnoses in the set of remaining diagnoses we can stop further query generation if one diagnosis is highly probable and all other remaining diagnoses are highly improbable.

**Example 2.** *Consider an ontology $O$ with the terminology $\mathcal{T}$:*

$$
\begin{array}{ll}
ax_1 : A_1 \sqsubseteq A_2 \sqcap M_1 \sqcap M_2 & ax_4 : M_2 \sqsubseteq \forall s.A \sqcap D \\
ax_2 : A_2 \sqsubseteq \neg\exists s.M_3 \sqcap \exists s.M_2 & ax_5 : M_3 \equiv B \sqcup C \\
ax_3 : M_1 \sqsubseteq \neg A \sqcap B &
\end{array}
$$

*and the background theory containing the assertions $\mathcal{A} : \{A_1(w), A_1(u), s(u, w)\}$.*

The ontology is inconsistent and includes two minimal conflict sets: $\{\langle ax_1, ax_3, ax_4\rangle, \langle ax_1, ax_2, ax_3, ax_5\rangle\}$. To restore consistency, the user should modify all axioms of at least one minimal diagnosis:

$$
\begin{array}{ll}
\mathcal{D}_1 : [ax_1] & \mathcal{D}_3 : [ax_4, ax_5] \\
\mathcal{D}_2 : [ax_3] & \mathcal{D}_4 : [ax_4, ax_2]
\end{array}
$$

Following the same approach as in the first example, we compute a set of possible queries and corresponding partitions using the algorithm presented above. A set of irreducible queries possible in Example 2 and their partitions are presented in Table 3. These queries partition the set of diagnoses **D** in a way that makes the application of myopic strategies, such as "Split-in-half", inefficient. A greedy algorithm based on such a heuristic would select the first query $Q_1$ as the next query, since there is no query that cuts the set of diagnoses in half. If $\mathcal{D}_4$ is the target diagnosis then $Q_1$ will be positively evaluated by an oracle (see Figure 1). On the next iteration the algorithm would also choose a suboptimal query since there is no partition that divides the diagnoses $\mathcal{D}_1$, $\mathcal{D}_2$, and $\mathcal{D}_4$ into two equal groups. Consequently, it selects the first untried query $Q_2$. The oracle answers positively, and the algorithm identifies query $Q_4$ to differentiate between $\mathcal{D}_1$ and $\mathcal{D}_4$.

However, in real-world settings the assumption that all axioms fail with the same probability is rarely the case. For example, Roussey et al. [5] present a list of "anti-patterns". Each anti-pattern is a set of axioms, like $\{C1 \sqsubseteq \forall R.C2, C1 \sqsubseteq \forall R.C3, C2 \equiv \neg C3\}$, that correspond to a minimal conflict set. The study performed by the authors shows that such conflict sets occur often

| Query | $\mathbf{D^P}$ | $\mathbf{D^N}$ | $\mathbf{D^\emptyset}$ |
|---|---|---|---|
| $Q_1 : \{B(w)\}$ | $\{\mathcal{D}_2, \mathcal{D}_3, \mathcal{D}_4\}$ | $\{\mathcal{D}_1\}$ | $\emptyset$ |
| $Q_2 : \{B(w), C(w)\}$ | $\{\mathcal{D}_3, \mathcal{D}_4\}$ | $\{\mathcal{D}_1, \mathcal{D}_2\}$ | $\emptyset$ |
| $Q_3 : \{B(w), C(w), Q(w)\}$ | $\{\mathcal{D}_4\}$ | $\{\mathcal{D}_1, \mathcal{D}_2, \mathcal{D}_3\}$ | $\emptyset$ |

Table 2: Possible queries in Example 1

| Query | $\mathbf{D^P}$ | $\mathbf{D^N}$ | $\mathbf{D^\emptyset}$ |
|---|---|---|---|
| $Q_1 : \{B \sqsubseteq M_3\}$ | $\{\mathcal{D}_1, \mathcal{D}_2, \mathcal{D}_4\}$ | $\{\mathcal{D}_3\}$ | $\emptyset$ |
| $Q_2 : \{B(w)\}$ | $\{\mathcal{D}_3, \mathcal{D}_4\}$ | $\{\mathcal{D}_2\}$ | $\{\mathcal{D}_1\}$ |
| $Q_3 : \{M_1 \sqsubseteq B\}$ | $\{\mathcal{D}_1, \mathcal{D}_3, \mathcal{D}_4\}$ | $\{\mathcal{D}_2\}$ | $\emptyset$ |
| $Q_4 : \{M_1(w), M_2(u)\}$ | $\{\mathcal{D}_2, \mathcal{D}_3, \mathcal{D}_4\}$ | $\{\mathcal{D}_1\}$ | $\emptyset$ |
| $Q_5 : \{A(w)\}$ | $\{\mathcal{D}_2\}$ | $\{\mathcal{D}_3, \mathcal{D}_4\}$ | $\{\mathcal{D}_1\}$ |
| $Q_6 : \{M_2 \sqsubseteq D\}$ | $\{\mathcal{D}_1, \mathcal{D}_2\}$ | $\emptyset$ | $\{\mathcal{D}_3, \mathcal{D}_4\}$ |
| $Q_7 : \{M_3(u)\}$ | $\{\mathcal{D}_4\}$ | $\emptyset$ | $\{\mathcal{D}_1, \mathcal{D}_2, \mathcal{D}_3\}$ |

Table 3: Possible queries in Example 2

in practice and therefore can be used to compute probabilities of diagnoses.

The approach that we follow in this paper was suggested by Rector et al. [4] and considers the syntax of a knowledge representation language, such as restrictions, conjunction, negation, etc., rather than axioms to describe a failure pattern. For instance, if a user frequently modifies the universal to the existential quantifier and vice versa in order to restore coherency, then we can assume that axioms including restrictions are more probable to fail than the other ones. In [4] the authors report that in most cases inconsistent ontologies were created because users (a) mix up $\forall r.S$ and $\exists r.S$, (b) mix up $\neg\exists r.S$ and $\exists r.\neg S$, (c) mix up $\sqcup$ and $\sqcap$, (d) wrongly assume that classes are disjoint by default or overuse disjointness, (e) wrongly apply negation. Observing that misuses of quantifiers are more likely than other failure patterns one might find that the axioms $ax_2$ and $ax_4$ are more likely to be faulty than $ax_3$ (because of the use of quantifiers), whereas $ax_3$ is more likely to be faulty than $ax_5$ and $ax_1$ (because of the use of negation).

Detailed justifications of diagnoses probabilities are given in the next section. However, let us assume some probability distribution of the faults according to the observations presented above such that: (a) the diagnosis $\mathcal{D}_2$ is the most probable one, i.e. single fault diagnosis of an axiom containing a negation; (b) although $\mathcal{D}_4$ is a double fault diagnosis, it follows $\mathcal{D}_2$ closely as its axioms contain quantifiers; (c) $\mathcal{D}_1$ and $\mathcal{D}_3$ are significantly less probable than $\mathcal{D}_1$ because conjunction/disjunction in $ax_1$ and $ax_5$ have a significantly lower fault probability than negation in $ax_3$. Taking into account this information it is almost useless to ask query $Q_1$ because it is highly probable that the target diagnosis is either $\mathcal{D}_2$ or $\mathcal{D}_4$ and, therefore, it is highly probable that the oracle will respond with *yes*. Instead, asking $Q_3$ is more informative because given any possible answer we can exclude one of the highly probable diagnoses, i.e. either $\mathcal{D}_2$ or $\mathcal{D}_4$. If the oracle responds to $Q_3$ with *no* then $\mathcal{D}_2$ is the only remaining diagnosis. However, if the oracle responds with *yes*, diagnoses $\mathcal{D}_4$, $\mathcal{D}_3$, and $\mathcal{D}_1$ remain, where $\mathcal{D}_4$ is significantly more probable compared to diagnoses $\mathcal{D}_3$ and $\mathcal{D}_1$. We can stop, if the difference between the probabilities of the diagnoses is high enough such that $\mathcal{D}_4$ can be accepted as the target diagnosis. Otherwise, additional questions may be required. This strategy can lead to a substantial reduction in the number of queries compared to myopic approaches as we will show in our evaluation.

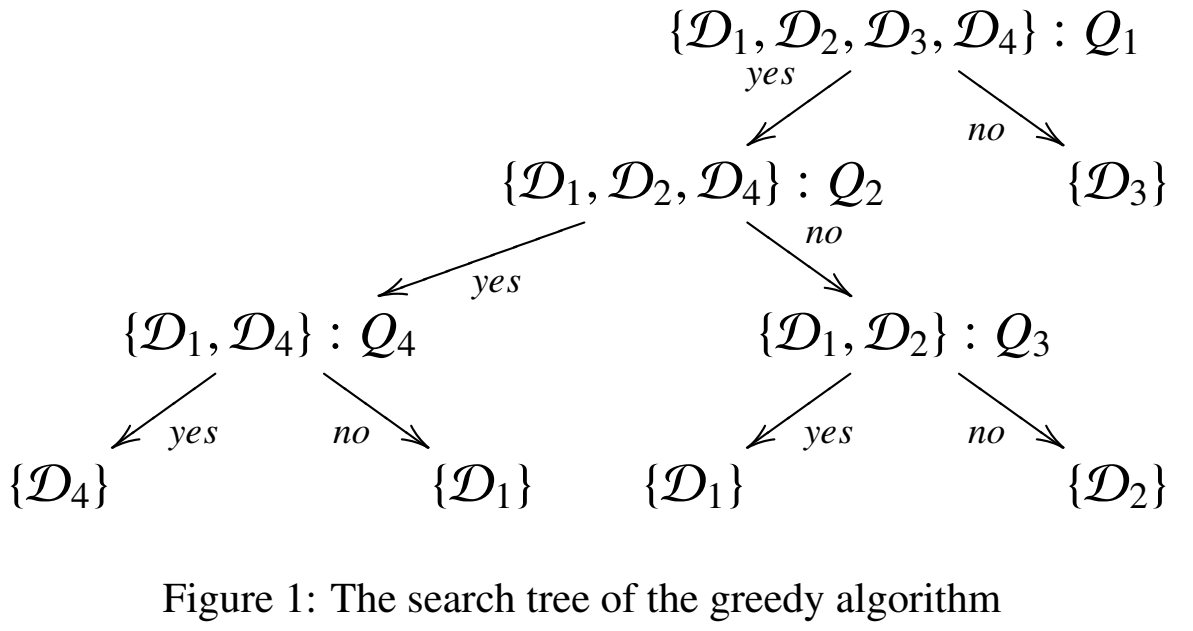

Figure 1: The search tree of the greedy algorithm

Note that in real-world application scenarios failure patterns and their probabilities can be discovered by analyzing actions of a user in an ontology editor, like Protégé, while debugging an ontology. In this case it is possible to "personalize" the measurement selection algorithm such that it will prefer user-specific faults. However, as our evaluation shows only a rough estimate of the probabilities is sufficient to outperform the "Split-in-half" heuristic.

## 3. Entropy-based query selection

To select the best query we make the assumption that knowledge is available about the a-priori failure probabilities. In our approach we follow the proposal of Rector et al. [4] and describe failure patterns employing the syntax of description logics or some other knowledge representation language, such as OWL. That is, either a user should express own beliefs in terms of the probability of a syntax element like $\forall, \exists, \sqcup$, etc. to be erroneous; or the debugger can compute these probabilities by analyzing how often a particular syntax element occurred in target diagnoses of different debugging sessions. If no information about failures is available then the debugger can initialize all probabilities with some small number.

Given failure probabilities of all syntax elements of a knowledge representation language we can compute the failure probability of an axiom

$$p(ax_i) = p(se_1 \cap se_2 \cap \cdots \cap se_n)$$

where $se_1 \dots se_n$ are syntax elements occurring in $ax_i$. Assuming that all syntax elements fail independently, i.e. an erroneous usage of a syntax element $se_i$ makes it neither more nor less probable that a syntax element $se_j$ is faulty, the failure probability of an axiom is defined as:

$$p(ax_i) = \sum_{1 \le i \le n} p(se_i) - \sum_{1 \le i < j \le n} p(se_i)p(se_j) + \dots + (-1)^{n-1} \prod_{1 \le i \le n} p(se_i) \quad (1)$$

For instance, the axiom $ax_2$ in Example 2 includes the following syntax elements $\{\sqsubseteq, \neg, \exists, \sqcup, \exists\}$. If among other failure probabilities the user provides that $p(\sqsubseteq) = 0.001, p(\neg) = 0.01, p(\exists) = 0.05$ and $p(\sqcup) = 0.001$ then $p(ax_2) = 0.108$.

Given the failure probabilities $p(ax_i)$ of axioms, the diagnosis algorithm first calculates the a-priori probability $p(\mathcal{D}_j)$ that $\mathcal{D}_j$ is the target diagnosis. Since all axioms fail independently, this probability can be computed as [15]:

$$p(\mathcal{D}_j) = \prod_{ax_n \in \mathcal{D}_j} p(ax_n) \prod_{ax_m \notin \mathcal{D}_j} 1 - p(ax_m) \quad (2)$$

The prior probabilities for diagnoses are then used to initialize an iterative algorithm that includes two main steps: (a) selection of the best query and (b) update of the diagnoses probabilities given the query feedback.

According to information theory the best query is the one that, given the answer of an oracle, minimizes the expected entropy of the set of diagnoses [15]. Let $p(Q_i = yes)$ be the probability that query $Q_i$ is answered with *yes* and $p(Q_i = no)$ be the probability for the answer *no*. Let $p(\mathcal{D}_j|Q_i = yes)$ be the probability of diagnosis $\mathcal{D}_j$ after the oracle answers *yes* and $p(\mathcal{D}_j|Q_i = no)$ be the probability for the answer *no*. The expected entropy after querying $Q_i$ is:

$$H_e(Q_i) = \sum_{v \in \{yes,no\}} p(Q_i = v) \times - \sum_{\mathcal{D}_j \in \mathbf{D}} p(\mathcal{D}_j|Q_i = v) \log_2 p(\mathcal{D}_j|Q_i = v)$$

The query which minimizes the expected entropy is the best one based on a one-step-look-ahead information theoretic measure. This formula can be simplified to the following score function [15] which we use to evaluate all available queries and select the one with the minimum score to maximize information gain:

$$sc(Q_i) = \sum_{v \in \{yes,no\}} [p(Q_i = v) \log_2 p(Q_i = v)] + p(\mathbf{D_i^\emptyset}) + 1 \quad (3)$$

where $\mathbf{D_i^\emptyset}$ is the set of diagnoses which do not make any predictions for the query $Q_i$. $p(\mathbf{D_i^\emptyset})$ is the total probability of the diagnoses that predict no value for the query $Q_i$. Since, for a query $Q_i$ the set of diagnoses $\mathbf{D}$ can be partitioned into the sets $\mathbf{D_i^P}$, $\mathbf{D_i^N}$ and $\mathbf{D_i^\emptyset}$, the probability that an oracle will answer a query $Q_i$ with either *yes* or *no* can be computed as:

$$\begin{aligned} p(Q_i = yes) &= p(\mathbf{D_i^P}) + p(\mathbf{D_i^\emptyset})/2 \\ p(Q_i = no) &= p(\mathbf{D_i^N}) + p(\mathbf{D_i^\emptyset})/2 \end{aligned} \quad (4)$$

Under the assumption that for each diagnosis of $\mathbf{D_i^\emptyset}$ *both outcomes are equally likely* the probability that the set of diagnoses $\mathbf{D_i^\emptyset}$ predicts either $Q_i = yes$ or $Q_i = no$ is $p(\mathbf{D_i^\emptyset})/2$.

Because of Definition 1 each diagnosis is a unique partition of all axioms of an ontology $O$ into correct and faulty, all diagnoses are mutually exclusive events. Therefore the probabilities of their sets can be calculated as:

$$p(\mathbf{S_i}) = \sum_{\mathcal{D}_j \in \mathbf{S_i}} p(\mathcal{D}_j)$$

where $\mathbf{S_i}$ corresponds to the sets $\mathbf{D_i^P}$, $\mathbf{D_i^N}$ and $\mathbf{D_i^\emptyset}$ respectively.

Given the feedback $v$ of an oracle to the selected query $Q_s$, i.e. $Q_s = v$, we have to update the probabilities of the diagnoses to take the new information into account. The update is made using Bayes' rule for each $\mathcal{D}_j \in \mathbf{D}$:

$$p(\mathcal{D}_j|Q_s = v) = \frac{p(Q_s = v|\mathcal{D}_j)p(\mathcal{D}_j)}{p(Q_s = v)} \quad (5)$$

where the denominator $p(Q_s = v)$ is known from the query selection step (Equation 4) and $p(\mathcal{D}_j)$ is either a prior probability (Equation 2) or is a probability calculated using Equation 5 after a previous iteration of the debugging algorithm. We assign $p(Q_s = v|\mathcal{D}_j)$ as follows:

$$p(Q_s = v|\mathcal{D}_j) = \begin{cases} 1, & \text{if } \mathcal{D}_j \text{ predicted } Q_s = v; \\ 0, & \text{if } \mathcal{D}_j \text{ is rejected by } Q_s = v; \\ \frac{1}{2}, & \text{if } \mathcal{D}_j \in \mathbf{D_s^\emptyset} \end{cases}$$

**Example 1 (continued)** Suppose that the debugger is not provided with any information about possible failures and therefore it is assumed that all syntax elements fail with the same probability 0.01 and therefore $p(ax_i) = 0.01$. Using Equation 2 we can calculate probabilities for each diagnosis. For instance, $\mathcal{D}_1$ suggests that only one axiom $ax_1$ should be modified by the user. Hence, we can calculate the probability of diagnosis $D_1$ as follows $p(\mathcal{D}_1) = p(ax_1)(1 - p(ax_2))(1 - p(ax_3))(1 - p(ax_4)) = 0.0097$. All other minimal diagnoses have the same probability, since every other minimal diagnosis suggests the modification of one axiom. To simplify the discussion we only consider minimal diagnoses for the query selection. Therefore, the prior probabilities of the diagnoses can be normalized to $p(\mathcal{D}_j) = p(\mathcal{D}_j)/\sum_{\mathcal{D}_j \in \mathbf{D}} p(\mathcal{D}_j)$ and are equal to 0.25.

Given the prior probabilities of the diagnoses and a set of queries (see Table 2) we evaluate the score function (Equation 3) for each query. E.g. for the first query $Q_1 : \{B(w)\}$ the probability $p(\mathbf{D}^\emptyset) = 0$ and the probabilities of both the positive and negative outcomes are: $p(Q_1 = 1) = p(\mathcal{D}_2) + p(\mathcal{D}_3) + p(\mathcal{D}_4) = 0.75$ and $p(Q_1 = 0) = p(\mathcal{D}_1) = 0.25$. Therefore the query score is $sc(Q_1) = 0.1887$.

The scores computed during the initial stage (see Table 4) suggest that $Q_2$ is the best query. Note, we include in Table 4 the minimized queries. Taking into account that $\mathcal{D}_1$ is the target diagnosis the oracle answers *no* to the query. The additional information obtained from the answer is then used to update the probabilities of diagnoses using the Equation 5. Since $\mathcal{D}_1$ and $\mathcal{D}_2$ predicted this answer, their probabilities are updated, $p(\mathcal{D}_1) = p(\mathcal{D}_2) = 1/p(Q_2 = 1) = 0.5$. The probabilities of diagnoses $\mathcal{D}_3$ and $\mathcal{D}_4$ which are rejected by the outcome are also updated, $p(\mathcal{D}_3) = p(\mathcal{D}_4) = 0$.

| Query | Initial score | $Q_2 = yes$ |
|---|---|---|
| $Q_1 : \{B(w)\}$ | 0.1887 | **0** |
| $Q_2 : \{C(w)\}$ | **0** | 1 |
| $Q_3 : \{Q(w)\}$ | 0.1887 | 1 |

Table 4: Expected scores for queries ($p(ax_i) = 0.01$)

| Query | Initial score |
|---|---|
| $Q_1 : \{B(w)\}$ | **0.250** |
| $Q_2 : \{C(w)\}$ | 0.408 |
| $Q_3 : \{Q(w)\}$ | 0.629 |

Table 5: Expected scores for queries ($p(ax_1) = 0.025$, $p(ax_2) = p(ax_3) = p(ax_4) = 0.01$)

On the next iteration the algorithm recomputes the scores using the updated probabilities. The results show that $Q_1$ is the best query. The other two queries $Q_2$ and $Q_3$ are irrelevant since no information will be gained if they are performed. Given the negative feedback of an oracle to $Q_1$, we update the probabilities $p(\mathcal{D}_1) = 1$ and $p(\mathcal{D}_2) = 0$. In this case the target diagnosis $\mathcal{D}_1$ was identified using the same number of steps as the split-in-half heuristic.

However, if the user specifies directly that the first axiom is more likely to fail, e.g. $p(ax_1) = 0.025$, then the first query will be $Q_1 : \{B(w)\}$ (see Table 5). The recalculation of the probabilities given the negative outcome $Q_1 = 0$ sets $p(\mathcal{D}_1) = 1$ and $p(\mathcal{D}_2) = p(\mathcal{D}_3) = p(\mathcal{D}_4) = 0$. Therefore the debugger identifies the target diagnosis only in one step.

**Example 2 (continued)** Suppose that in $ax_4$ the user specified $\forall s.A$ instead of $\exists s.A$ and $\neg\exists s.M_3$ instead of $\exists s.\neg M_3$ in $ax_2$. Therefore $\mathcal{D}_4$ is the target diagnosis. Moreover, the debugger is provided with observations of three types of faults: (1) conjunction/disjunction occurs with probability $p_1 = 0.001$, (2) negation $p_2 = 0.01$, and (3) restrictions $p_3 = 0.05$. Using Equation 1 we can calculate the probability of the axioms containing an error: $p(ax_1) = 0.0019$, $p(ax_2) = 0.1074$, $p(ax_3) = 0.012$, $p(ax_4) = 0.051$, and $p(ax_5) = 0.001$. These probabilities are exploited to calculate the prior probabilities of the diagnoses (see Table 6) and to initialize the query selection process. To simplify matters we focus on the set of minimal diagnoses.

On the first iteration the algorithm determines that $Q_3$ is the best query and asks an oracle whether $O_t \models \{M_1 \sqsubseteq B\}$ is true or not (see Table 7). The obtained in-formation is then used to recalculate the probabilities of the diagnoses and to compute the next best query, i.e. $Q_4$, and so on. The query process stops after the third query, since $\mathcal{D}_4$ is the only diagnosis that has the probability $p(\mathcal{D}_4) > 0$.

Given the feedback of the oracle $Q_4 = yes$ for the second query, the updated probabilities of the diagnoses show that the target diagnosis has a probability of $p(\mathcal{D}_4) = 0.9918$ whereas $p(\mathcal{D}_3)$ is only 0.0082. In order to reduce the number of queries a user can specify a threshold, e.g. $\sigma = 0.95$. If the absolute difference in probabilities of two most probable diagnoses is greater than this threshold, the query process stops and returns the most probable diagnosis. Therefore, in this example the debugger based on the entropy query selection requires less queries than the "Split-in-half" heuristic. Note that already after the first answer $Q_3 = yes$ the most probable diagnosis $\mathcal{D}_4$ is three times more likely than the second most probable diagnosis $\mathcal{D}_1$. Given such a great difference we could suggest to stop the query process after the first answer by setting $\sigma = 0.65$.

## 4. Implementation details

The iterative ontology debugger (Algorithm 1) takes a faulty ontology $\mathcal{O}$ as input. Optionally, a user can provide a set of axioms $\mathcal{B}$ that are known to be correct as well as a set $P$ of axioms that must be entailed by the target ontology and a set $N$ of axioms that must not. If these sets are not given, the corresponding input arguments are initialized with $\emptyset$. Moreover, the algorithm takes a set $FP$ of fault probabilities for axioms $ax_i \in \mathcal{O}$, which can be computed as described in Section 3 by exploiting knowledge about typical user errors. The two other arguments $\sigma$ and $n$ are used to speed up the performance of the algorithm. $\sigma$ sets the diagnosis acceptance threshold that defines the absolute difference in probabilities of the two most probable diagnoses. The parameter $n$ defines a maximum number of most probable diagnoses that should be considered by the algorithm on each iteration. A further performance gain in Algorithm 1 can be achieved if we approximate the set of the $n$ most probable diagnoses with the set of the $n$ most probable *minimal* diagnoses, i.e. we neglect non-minimal diagnoses. We call this set of at most $n$ most probable minimal diagnoses the *leading diagnoses*. Note, under a reasonable assumption that the fault probability of each axiom $p(ax_i)$ is less than 0.5, it is the case that for every non-minimal diagnosis $ND$ a minimal diagnosis $\mathcal{D} \subset ND$ exists, which from Equation 2 is more probable than $ND$. Consequently the query selection algorithm operates on the set of minimal diagnoses instead of all diagnoses (including non-minimal ones). However, the algorithm can be adapted with moderate effort to consider non-minimal diagnoses.

We implemented the computation of diagnoses following the approach proposed by Friedrich et al. [8]. The authors employ the combination of two algorithms, QuickXplain [17] and HS-Tree [14]. In a standard implementation the latter is a breadth-first search algorithm that takes an ontology $\mathcal{O}$, sets of logical sentences $P$ and $N$, and the maximal number of most probable minimal diagnoses $n$ as an input. In particular, minimal hitting set generation and the search for minimal conflict sets is interleaved. This is motivated by the fact that for the generation of a subset of the set of all minimal diagnoses possibly only a subset of the set of all minimal conflict sets is needed. In our case we compute at most $n$ minimal diagnoses. This is an important property because the number of minimal conflict sets can grow exponential in the size of the ontology. Note, a minimal diagnosis is a minimal hitting set of all minimal conflict sets. However, in order to verify that a set of axioms is a minimal diagnosis, the set of all minimal conflict sets is not needed. In our implementation of HS-Tree we use the uniform-cost search strategy. Given additional information in terms of fault axiom probabilities $FP$, the algorithm expands a leaf node in a search-tree if it is an element of the maximum probability hitting set, given the currently found set of minimal conflict sets. The probability of each hitting set can computed using Equation 2. Consequently, the algorithm computes a set of diagnoses ordered by their probability starting from the most probable one. HS-Tree terminates if either the $n$ most probable minimal diagnoses are identified or there are no further minimal diagnoses.

The search algorithm computes minimal conflicts using QuickXplain. This algorithm, given a set of axioms $AX$ and a set of correct axioms $\mathcal{B}$ returns a minimal conflict set $CS \subseteq AX$, or $\emptyset$ if axioms $AX \cup \mathcal{B}$ are consistent. Minimal conflicts are computed on-demand by HS-Tree while exploring the search space.

In order to take past answers into account the HS-Tree updates the prior probabilities of the diagnoses by evaluating Equation 5. The query history is stored in $QH$ as well as in the updates of $P$, and $N$. As a result HS-Tree returns a set of tuples $\langle \mathcal{D}_i, p(\mathcal{D}_i) \rangle$ where $\mathcal{D}_i$ is contained in the set of the $n$ most probable minimal diagnoses (leading diagnoses) and $p(\mathcal{D}_i)$ is its probability using Equation 2 and Equation 5.

In the query-selection phase Algorithm 1 calls selectQuery function (Algorithm 2) to generate a tuple

| Answers | $\mathcal{D}_1$ | $\mathcal{D}_2$ | $\mathcal{D}_3$ | $\mathcal{D}_4$ |
|---|---|---|---|---|
| Prior | 0.0970 | 0.5874 | 0.0026 | 0.3130 |
| $Q_3 = yes$ | 0.2352 | 0 | 0.0063 | 0.7585 |
| $Q_3 = yes$, $Q_4 = yes$ | 0 | 0 | 0.0082 | 0.9918 |
| $Q_3 = yes$, $Q_4 = yes$, $Q_1 = yes$ | 0 | 0 | 0 | 1 |

Table 6: Probabilities of diagnoses after answers

| Queries | Initial | $Q_3 = yes$ | $Q_3 = yes$, $Q_4 = yes$ |
|---|---|---|---|
| $Q_1 : \{B \sqsubseteq M_3\}$ | 0.974 | 0.945 | **0.931** |
| $Q_2 : \{B(w)\}$ | 0.151 | 0.713 | 1 |
| $Q_3 : \{M_1 \sqsubseteq B\}$ | **0.022** | 1 | 1 |
| $Q_4 : \{M_1(w), M_2(u)\}$ | 0.540 | **0.213** | 1 |
| $Q_5 : \{A(w)\}$ | 0.151 | 0.713 | 1 |
| $Q_6 : \{M_2 \sqsubseteq D\}$ | 0.686 | 0.805 | 1 |
| $Q_7 : \{M_3(u)\}$ | 0.759 | 0.710 | 0.970 |

Table 7: Expected scores for queries

**Algorithm 1:** ONTODEBUGGING($\mathcal{O}, \mathcal{B}, P, N, FP, n, \sigma$)

**Input**: ontology $\mathcal{O}$, set of background axioms $\mathcal{B}$,
$P, N$ sets of sentences to be (not) entailed,
set of fault probabilities for axioms $FP$,
maximum number of most probable
minimal diagnoses $n$,
acceptance threshold $\sigma$

**Output**: a diagnosis $\mathcal{D}$

$DP \leftarrow \emptyset$; $QH \leftarrow \emptyset$; $T \leftarrow \langle\emptyset, \emptyset, \emptyset, \emptyset\rangle$;
**while** BELOWTHRESHOLD($DP, \sigma$) $\wedge$ GETSCORE($T$) $\neq 1$ **do**
  $DP \leftarrow$ HS-TREE($\mathcal{O}, \mathcal{B} \cup P, N, FP, QH, n$);
  $T \leftarrow$ SELECTQUERY($DP, \mathcal{O}, \mathcal{B}, P$);
  $Q \leftarrow$ GETQUERY($T$);
  **if** $Q = \emptyset$ **then exit loop**;
  **if** GETANSWER($\mathcal{O}_t \models Q$) **then** $P \leftarrow P \cup Q$;
  **else** $N \leftarrow N \cup Q$;
  $QH \leftarrow QH \cup \{T\}$;
**return** MOSTPROBABLEDIAGNOSIS($DP$);

$T = \left\langle Q, \mathbf{D^P}, \mathbf{D^N}, \mathbf{D^\emptyset} \right\rangle$, where $Q$ corresponds to the minimal score query (Equation 3) for the sets of diagnoses $\mathbf{D^P}, \mathbf{D^N}$ and $\mathbf{D^\emptyset}$. The generation algorithm implements a depth-first search as it removes the top element of the set $DP$ and calls itself recursively to generate all possible subsets of the leading diagnoses. In each leaf node of the search tree the GENERATE function calls CREATEQUERY to create a query given a set of diagnoses $\mathbf{D^P}$ as described in Section 2.2, i.e. computation of common entailments followed by a partitioning of the diagnoses. If a query for the set $\mathbf{D^P}$ does not exists or $\mathbf{D^P} = \emptyset$ then CREATEQUERY returns an empty tuple $T = \langle\emptyset, \emptyset, \emptyset, \emptyset\rangle$. In all other nodes of the tree the algorithm selects a tuple that corresponds to a query with the minimal score by using the GETSCORE function. The latter might implement the entropy-based measure (Equation 3), "Split-in-half" or any other preference criteria. Given an empty tuple $T = \langle\emptyset, \emptyset, \emptyset, \emptyset\rangle$ the function should return the highest possible score of 1. Moreover, if the scores are equal then the algorithm returns a tuple where $Q$ has the smallest cardinality in order to reduce the answering effort. By the function MINIMIZEQUERY the query $Q$ of the resulting tuple $\left\langle Q, \mathbf{D^P}, \mathbf{D^N}, \mathbf{D^\emptyset} \right\rangle$ is iteratively reduced by applying QUICKXPLAIN such that sets $\mathbf{D^P}$, $\mathbf{D^N}$ and $\mathbf{D^\emptyset}$ are preserved. However, MINIMIZEQUERY checks if the query was already minimzed.

In Algorithm 1 the function GETQUERY simply selects the query from the tuple stored in $T$ and subsequently the user is asked by GETANSWER. Depending on the answer of the oracle, Algorithm 1 extends either the set $P$ or the set $N$. This is done to exclude corresponding diagnoses from the results of HS-TREE in further iterations. Note, the algorithm can be easily adapted to allow the oracle to reject a query if the answer is unknown. In this case the algorithm proceeds with the next best query until no further queries are available.

Algorithm 1 stops if there is a diagnosis probability above the acceptance threshold $\sigma$ or if no query can be used to differentiate between the remaining diagnoses (i.e. the score of the minimal score query is 1). The most probable diagnosis is then returned to the user. If it is impossible to differentiate between a number of highly probable minimal diagnoses, the algorithm returns a set

**Algorithm 2:** selectQuery($DP, O, \mathcal{B}, P$)

**Input**: $DP$ if set of pairs $\langle \mathcal{D}_i, p(\mathcal{D}_i)\rangle$, $O$ ontology, $\mathcal{B}$ set of background axioms, $P$ set of axioms that must be entailed by the target ontology

**Output**: a tuple $\left\langle Q, \mathbf{D^P}, \mathbf{D^N}, \mathbf{D^\emptyset}\right\rangle$

$T \leftarrow$ generate($\emptyset, DP, O, \mathcal{B}, P$);<br>
**return** minimizeQuery($T$);

**function** generate ($\mathbf{D^P}, DP, O, \mathcal{B}, P$) **returns** a tuple $\left\langle Q, \mathbf{D^P}, \mathbf{D^N}, \mathbf{D^\emptyset}\right\rangle$<br>
  **if** $DP = \emptyset$ **then**<br>
    **return** createQuery ($\mathbf{D^P}, O, \mathcal{B}, P$);<br>
  $\langle \mathcal{D}, p(\mathcal{D})\rangle \leftarrow$ pop ($DP$);<br>
  *left* $\leftarrow$ generate ($\mathbf{D^P}, DP, O, \mathcal{B}, P$);<br>
  *right* $\leftarrow$ generate ($\mathbf{D^P} \cup \{\langle \mathcal{D}, p(\mathcal{D})\rangle\}$, $DP, O, \mathcal{B}, P$);<br>
  **if** getScore *(left)* < getScore *(right)* **then**<br>
    **return** *left*;<br>
  **else if** getScore *(left)* > getScore *(right)* **then return** *right*;<br>
  *left* $\leftarrow$ minimizeQuery(*left*);<br>
  *right* $\leftarrow$ minimizeQuery(*right*);<br>
  **return** minCardinalityQuery (*left*, *right*);

that includes all of them. Moreover, in the first case (termination on $\sigma$), the algorithm can continue, if the user is not satisfied with the returned diagnosis and at least one query exists.

Additional performance improvements can be achieved by using greedy strategies in Algorithm 2. The idea is to guide the search in a way that a leaf node of the left-most branch of a search tree will contain such a set of diagnoses $\mathbf{D^P}$ that might result in a tuple $\left\langle Q, \mathbf{D^P}, \mathbf{D^N}, \mathbf{D^\emptyset}\right\rangle$ with a low-score query. This method is based on the property of Equation 3 that $sc(Q) = 0$ if

$$\sum_{\mathcal{D}_i \in \mathbf{D^P}} p(\mathcal{D}_i) = \sum_{\mathcal{D}_j \in \mathbf{D^N}} p(\mathcal{D}_j) = 0.5 \quad \text{and} \quad p(\mathbf{D^\emptyset}) = 0$$

Consequently, the query selection problem can be presented as a two-way number partitioning problem: Given a set of numbers, divide them into two sets such that the difference between the sums of the numbers in each set is as small as possible. The Complete Karmarkar-Karp (CKK) algorithm [18], which is one of the best algorithms developed for the two-way partitioning problem, corresponds to an extension of the Algorithm 2 with a set differencing heuristic [19]. The algorithm stops if either the optimal solution to the two-way partitioning problem is found or there are no further subsets to be investigated.

The main drawback of CKK applied to the query selection is that none of the pruning techniques can be used, since we cannot guarantee that a query can always be generated for a given set of diagnoses $\mathbf{D^P}$. Even if the algorithm finds an optimal solution to the two-way partitioning problem, it still has to investigate all subsets of the set of diagnoses in order to find the minimum score query. To avoid this exhaustive search we extended CKK with one more termination criteria. Namely, the search stops if a query is found with a score below some predefined threshold $\gamma$.

## 5. Evaluation

The evaluation of our approach was performed using generated examples and real-world ontologies presented in Table 8. We employed generated examples to perform controlled experiments where the number of minimal diagnoses and their cardinality could be varied to make the identification of the target diagnosis more difficult. The main goal of the experiment using real-world ontologies is to demonstrate the applicability of our approach in real-world settings.

For the first test we created a generator which takes a consistent and coherent ontology, a set of fault patterns together with their probabilities, the minimum number of minimum cardinality diagnoses $m$, and the required cardinality $|\mathcal{D}_t|$ of these minimum cardinality diagnoses as inputs. For the tests we assume that the target diagnosis has cardinality $|\mathcal{D}_t|$. The output of the generator is an alteration of the input ontology for which at least the given number of minimum cardinality diagnoses with the required cardinality exist. In order to introduce inconsistencies and incoherences, the generator applies fault patterns randomly to the input ontology depending on their probabilities.

In this experiment we took five fault patterns from a case study reported by Rector et al. [4] and assigned fault probabilities according to their observations of typical user errors. Thus we assumed that in the cases (a) and (b) (see Section 2.2), when an axiom includes some roles (i.e. property assertions), axiom descriptions are faulty with a probability of 0.025, in the cases (c) and (d) 0.01 and in the case (e) 0.001. In each iteration the generator randomly selected an axiom to be altered and applied a fault pattern to this axiom. Next, another axiom was selected using the concept taxonomy and altered correspondingly to introduce an incoherency/inconsistency. The fault patterns were ran-domly selected in each step using the probabilities provided above.

| | Ontology | Axioms | #C/#P/#I | #CS/min/max | #D/min/max | Domain |
|---|---|---|---|---|---|---|
| 1. | Chemical | 114 | 48/20/0 | 6/5/6 | 6/1/3 | Chemical elements |
| 2. | Koala | 44 | 21/5/6 | 3/4/4 | 10/1/3 | Training |
| 3. | Sweet-JPL | 2579 | 1537/121/50 | 8/1/13 | 13/8/8 | Earthscience |
| 4. | miniTambis | 173 | 183/44/0 | 3/3/6 | 48/3/3 | Biological science |
| 5. | University | 50 | 30/12/4 | 4/3/5 | 90/3/4 | Training |
| 6. | Economy | 1781 | 339/53/482 | 8/3/4 | 864/4/9 | Mid-level |
| 7. | Transportation | 1300 | 445/93/183 | 9/2/6 | 1782/6/9 | Mid-level |

Table 8: Dianosis results for some real-world ontologies presented in [7]. #C/#P/#I are the numbers of concepts, properties, and individuals in an ontology. #CS/min/max are the number of conflict sets, their minimum and maximum cardinality. The same notation is used for diagnoses #D/min/max. These ontologies are available upon request.

For instance, given the description of a randomly selected concept $A$ and the fault pattern "misuse of negation", we added the construct $\sqcap\neg X$ to the description of $A$, where $X$ is a new concept name. Next, we randomly selected concepts $B$ and $S$ such that $S \sqsubseteq A$ and $S \sqsubseteq B$ and added $\sqcap X$ to the description of $B$. During the generation process, we applied the HS-Tree algorithm after each introduction of an incoherency/inconsistency to control two parameters: the minimum number of minimal cardinality diagnoses in the ontology and their cardinality. The generator continues to introduce incoherences/inconsistencies until the specified parameter values are reached. For instance, if the minimum number of minimum cardinality diagnoses is equal to $m = 6$ and their cardinality is $|\mathcal{D}_t| = 4$, then the generated ontology will include at least 6 diagnoses of cardinality 4 and possibly some additional number of minimal diagnoses of higher cardinalities.

The resulting faulty ontology as well as the fault patterns and their probabilities were inputs for the ontology debugger. The acceptance threshold $\sigma$ was set to 0.95 and the number of most probable minimal diagnoses $n$ was set to 9. One of the minimal diagnoses with the required cardinality was randomly selected as the target diagnosis. Note, the target ontology is not equal to the original ontology, but rather is a corrected version of the altered one, in which the faulty axioms were repaired by replacing them with their original (correct) versions according to the target diagnosis. The tests were done on ontologies bike2 to bike9, bcs3, galen and galen2 from Racer's benchmark suite[2].

The average results of the evaluation performed on each test suite (presented in Figure 2) show that the entropy-based approach outperforms the "Split-in-half" heuristic as well as the random query selection strategy by more than 50% for the $|\mathcal{D}_t| = 2$ case due to its ability to estimate the probabilities of diagnoses and to stop when the target diagnosis crossed the acceptance threshold. On average the algorithm required 8 seconds to generate a query. Figure 2 also shows that the number of required queries increases as the cardinality of the target diagnosis increases. This holds for the random and "Split-in-half" methods (not depicted) as well. However, the entropy-based approach is still better than the "Split-in-half" method even for diagnoses with increasing cardinality. The approach required more queries to discriminate between high cardinality diagnoses because the prior probabilities of these diagnoses tend to converge.

In the tests performed on the real-world ontologies we evaluated the performance of the entropy-based debugging algorithm given different user estimations of prior fault probabilities. The priors are very important since they are used by the entropy-based method to identify the best query to be asked. Given some misleading priors the entropy-based algorithm might require more queries to identify the target diagnosis. In our experiment we differentiated between three different distributions of the prior fault probabilities: extreme, moderate and uniform (see Figure 3 for an example). The *extreme distribution* simulates a situation when a user assigns very hight failure probabilities to a small number of syntax elements. That is, the user is quite sure that exactly these elements are causing a fault. For instance, the user has problems with formulating restrictions in OWL whereas all other elements, like subsumption, conjunction, etc., used in a faulty ontology are well understood. In the case of a *moderate distribution* the user provides a slight bias towards some syntax elements. This distribution has the same motivation as the extreme, however, in this case the user is less sure about possible causes of the problem. Both extreme and moderate distributions correspond to the exponential distri-bution with $\lambda = 1.75$ and $\lambda = 0.5$ respectively. The *uniform distribution* models the situation when the user did not provide any prior fault probabilities and the system assigns equal probabilities to all syntax elements found in a faulty ontology. Of course the user can make a mistake while estimating the priors and provide higher fault probabilities to elements that are correct. Therefore, for each of the three distributions we differentiate between good, average and bad cases. In the *good case* the user's estimates of the prior fault probabilities are correct and the target diagnosis receives a high probability. The *average case* corresponds to the situation when the target diagnosis is neither favored nor penalized by the priors. In the *bad case* the prior distribution predicts the target diagnosis incorrectly and, consequently, its probability is quite low.

[2] Available at `http://www.racer-systems.com/products/download/benchmark.phtml`

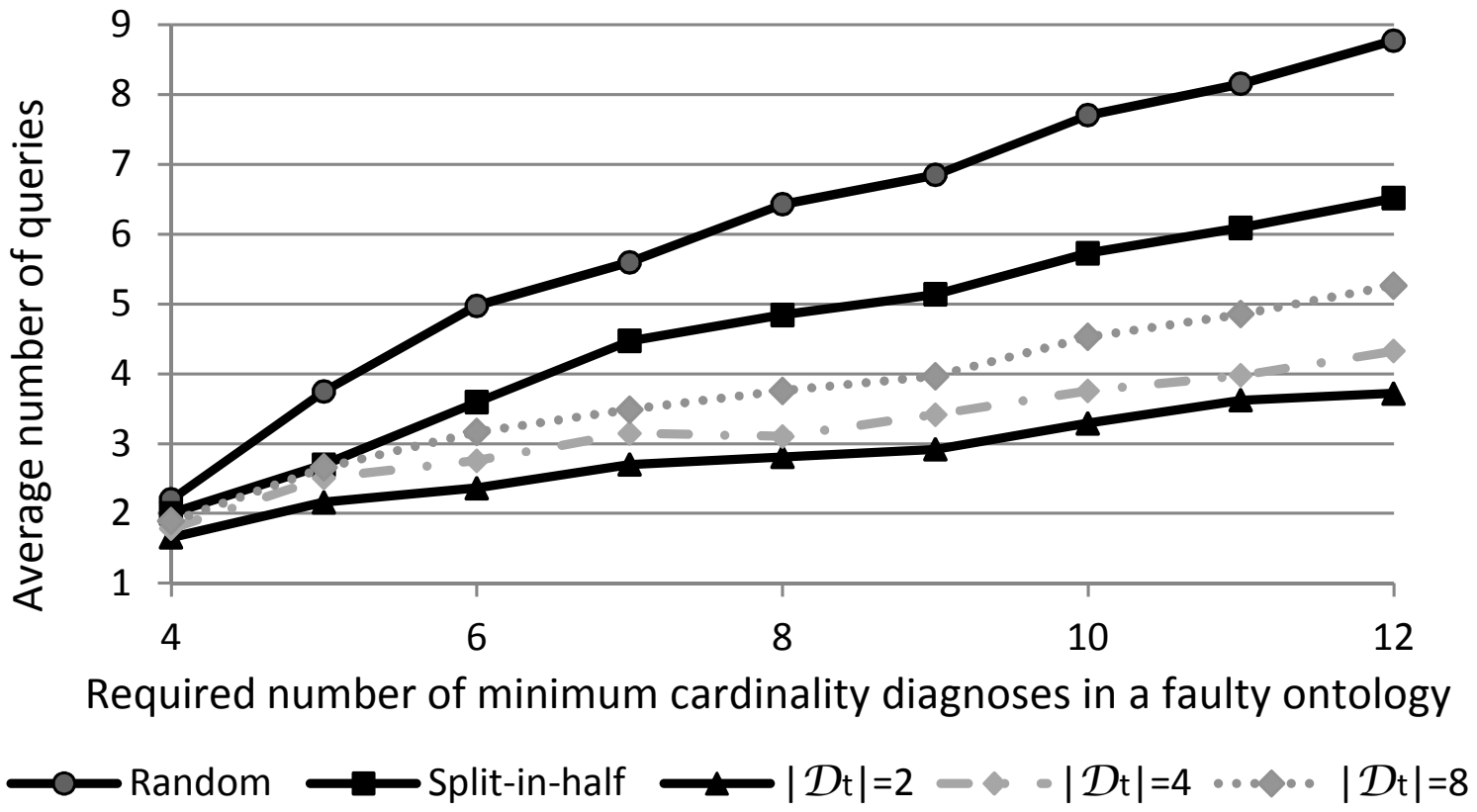

Figure 2: Average number of queries required to select the target diagnosis $\mathcal{D}_t$ with threshold $\sigma = 0.95$. Random and "Split-in-half" are shown for the cardinality of minimal diagnoses $|\mathcal{D}_t| = 2$.

We executed 30 tests for each of the combinations of the distributions and cases with acceptance threshold $\sigma = 0.85$ and number of most probable minimal diagnoses $n = 9$. Each iteration started with the generation of a set of prior fault probabilities of syntax elements by sampling from a selected distribution (extreme, moderate or uniform). Given the priors we computed the set of all minimal diagnoses $\mathbf{D}$ of a given ontology and selected the target one according to the chosen case (good, average or bad). In the good case the prior probabilities favor the target diagnosis and, therefore, it should be selected from the diagnoses with high probability. The set of diagnoses was ordered according to their probabilities and the algorithm iterated through the set starting from the most probable element. In each iteration $j$ a diagnosis $\mathcal{D}_j$ was added to the set $G$ if $\sum_{i \leq j} p(\mathcal{D}_i) \leq \frac{1}{3}$ and to the set $A$ if $\sum_{i \leq j} p(\mathcal{D}_i) \leq \frac{2}{3}$. The obtained set $G$ contained all most probable diagnoses which we considered as good. All diagnoses in the set $A \setminus G$ were classified as average and the remaining diagnoses $\mathbf{D} \setminus A$ as bad. Depending on the selected case we randomly selected one of the diagnoses as the target from the appropriate set.

The results of the evaluation presented in Table 9 show that the entropy-based query selection approach clearly outperforms "Split-in-half" in good and average cases for the three probability distributions. The plot of average number of queries required to identify the target diagnosis presented in Figure 4 shows that the performance of the entropy-based method does not depend on the type of the distribution provided by the user. In the uniform case the better results were observed since the diagnoses have different cardinality and structure, i.e. they include different syntax elements. Consequently, even if equal probabilities for all syntax elements (uniform distribution) are given, the probabilities of diagnoses are different. These differences provided enough bias to the entropy-based method. Only in the case of Sweet-JPL ontology the bias was insufficient and sometimes misleading since all diagnoses in this ontology are of the same cardinality and have similar structure. The major loss of performance can only be observed if the user provided misleading priors making the target diagnosis improbable. Therefore, we can conclude that the user should provide only some rough estimates of the prior fault probabilities that, however, favor the target diagnosis. The differences between probabilities of individual syntax elements are not influencing the results of the query selection and effect only the number of outliers, i.e. the cases when the diagnosis approach required either few or many queries compared to the average.

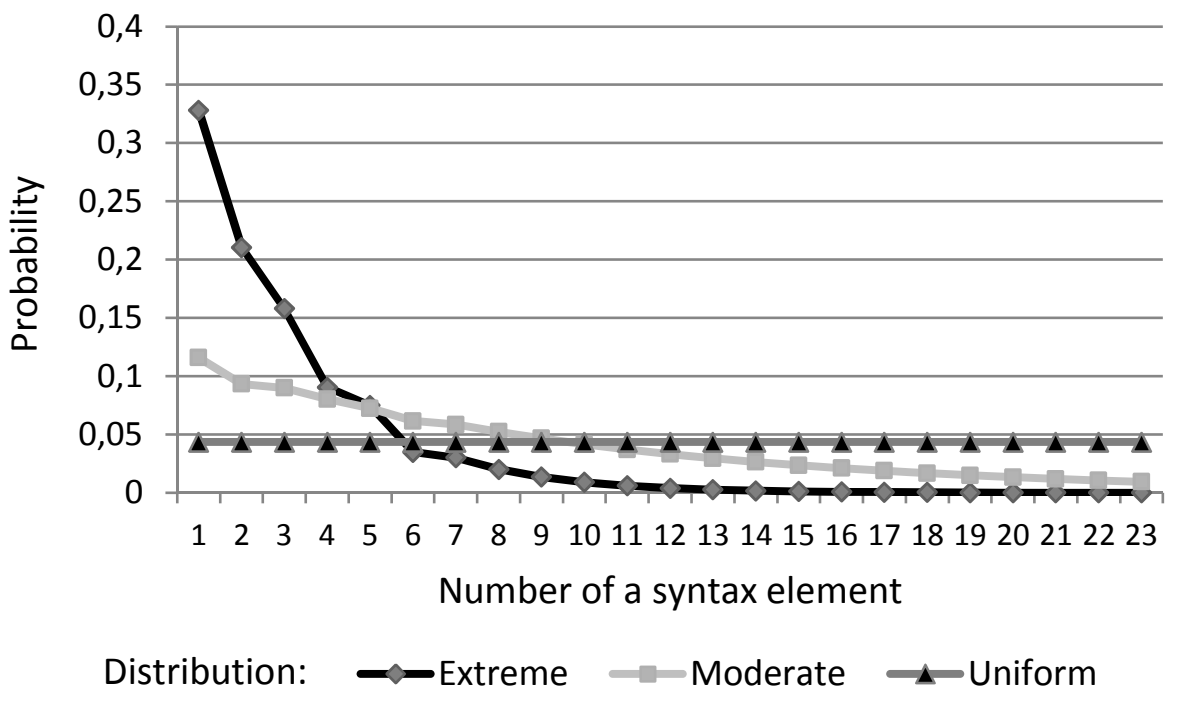

Figure 3: Example of prior fault probabilities of syntax elements sampled from extreme, moderate and uniform distributions.

Note that "Split-in-half" is inefficient in comparison to the entropy method in all good, average and bad cases when applied to the ontologies with a big number of diagnoses, such as Economy and Transportation. The main problem is that no stop criteria can be used with the greedy method as it does not provide any ordering on the set of diagnoses. Therefore, it has to continue until no further queries can be generated, i.e. only one minimal diagnosis exists or their are no discriminating queries.

Another interesting observation is that often both methods eliminated a bigger than $n$ number of diagnoses in one iteration. For instance, in the case of Transportation ontology both methods were able to remove hundreds of minimal diagnoses with a small number of queries. The main reason for this behavior are relations between the diagnoses. That is, addition of a query to either $P$ or $N$ allows the method to remove not only the diagnoses in sets $\mathbf{D^P}$ or $\mathbf{D^N}$, but also some unobserved diagnoses, that were not in any of the sets of $n$ leading diagnoses computed by HS-Tree. Given the sets $P$ and $N$ HS-Tree automatically invalidates all diagnoses, which do not fulfill the requirements (see Definition 1).

The extended CKK method presented in Section 4 was evaluated in the same settings as the complete Algorithm 2 with acceptance threshold $\gamma = 0.1$. The obtained results presented in Figure 5 show that the extended CKK method improves the time of a debugging session by at least 50% while requiring on average 0.2 queries more than Algorithm 2. In some cases (mostly for the uniform distribution) the debugger using greedy search required even less queries than Algorithm 2 because of the inherent uncertainty of the domain.

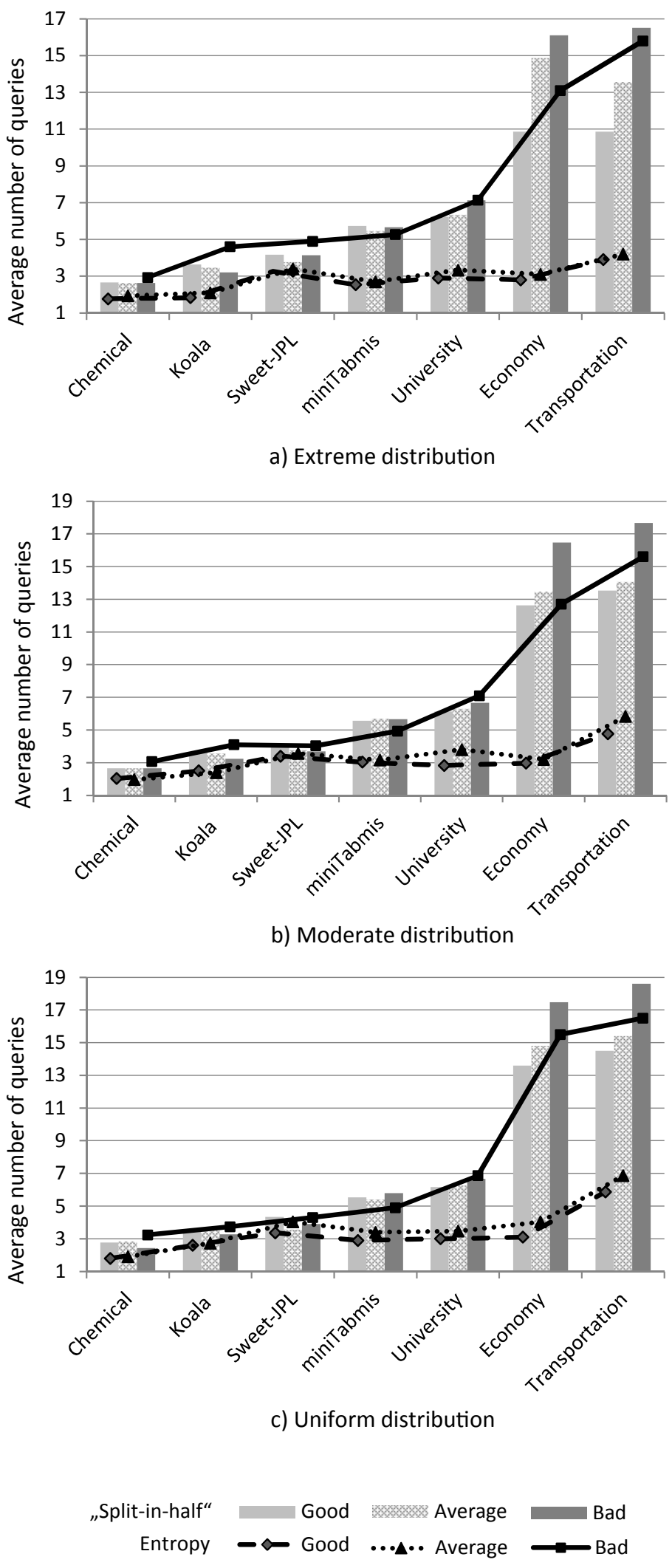

Figure 4: Average number of queries required to identify the target diagnosis.

## 6. Related work

To the best of our knowledge no sequential ontology debugging methods (neither employing "Split-in-half" nor entropy-based methods) have been proposed to debug faulty ontologies so far. Diagnosis methods for ontologies are introduced in [6, 7, 8]. Ranking of diagnoses and proposing a target diagnosis is presented in [10]. This method uses a number of measures such as: (a) the frequency with which an axiom appears in conflict sets, (b) impact on an ontology in terms of its "lost" entailments when some axiom is modified or removed, (c) ranking of test cases, (d) provenance information about the axiom, and (e) syntactic relevance.

Entropy-based query selection

| Ontology | Case | Distribution | | | | | | | | |
|---|---|---|---|---|---|---|---|---|---|---|
| | | Extreme | | | Moderate | | | Uniform | | |
| | | min | avg | max | min | avg | max | min | avg | max |
| Chemical | Good | 1 | 1.77 | 3 | 1 | 2.03 | 3 | 1 | 1.8 | 3 |
| | Avg. | 1 | 1.93 | 3 | 1 | 1.97 | 3 | 1 | 1.9 | 3 |
| | Bad | 2 | 2.93 | 4 | 2 | 3.07 | 4 | 2 | 3.23 | 4 |
| Koala | Good | 1 | 1.83 | 3 | 1 | 2.5 | 4 | 2 | 2.6 | 3 |
| | Avg. | 1 | 2.1 | 4 | 1 | 2.4 | 4 | 2 | 2.73 | 3 |
| | Bad | 2 | 4.6 | 8 | 2 | 4.1 | 6 | 3 | 3.73 | 5 |
| Sweet-JPL | Good | 1 | 3.27 | 6 | 1 | 3.4 | 5 | 3 | 3.37 | 4 |
| | Avg. | 1 | 3.4 | 6 | 1 | 3.57 | 5 | 3 | 4.03 | 5 |
| | Bad | 3 | 4.9 | 7 | 3 | 4.03 | 7 | 3 | 4.3 | 6 |
| miniTambis | Good | 1 | 2.53 | 4 | 2 | 3.03 | 4 | 2 | 2.9 | 3 |
| | Avg. | 1 | 2.7 | 4 | 2 | 3.17 | 4 | 3 | 3.4 | 4 |
| | Bad | 3 | 5.27 | 9 | 3 | 4.93 | 8 | 3 | 4.9 | 7 |
| University | Good | 2 | 2.9 | 4 | 2 | 2.83 | 3 | 3 | 3 | 3 |
| | Avg. | 1 | 3.33 | 7 | 3 | 3.8 | 5 | 3 | 3.47 | 5 |
| | Bad | 4 | 7.13 | 22 | 3 | 7.1 | 13 | 4 | 6.87 | 10 |
| Economy | Good | 2 | 2.8 | 3 | 2 | 2.96 | 3 | 3 | 3.1 | 5 |
| | Avg. | 2 | 3.1 | 4 | 3 | 3.2 | 4 | 3 | 4.03 | 5 |
| | Bad | 8 | 13.1 | 20 | 4 | 12.7 | 21 | 8 | 15.5 | 20 |
| Transportation | Good | 3 | 3.9 | 6 | 3 | 4.76 | 8 | 3 | 5.86 | 8 |
| | Avg. | 3 | 4.2 | 6 | 3 | 5.83 | 9 | 3 | 6.8 | 9 |
| | Bad | 9 | 15.8 | 31 | 10 | 15.6 | 20 | 8 | 16.5 | 30 |

"Split-in-half" query selection

| | | | | | | | | | | |
|---|---|---|---|---|---|---|---|---|---|---|
| Chemical | Good | 2 | 2.67 | 3 | 2 | 2.67 | 3 | 2 | 2.77 | 3 |
| | Avg. | 2 | 2.63 | 3 | 2 | 2.67 | 3 | 2 | 2.83 | 3 |
| | Bad | 2 | 2.63 | 3 | 2 | 2.67 | 3 | 2 | 2.43 | 3 |
| Koala | Good | 3 | 3.63 | 4 | 2 | 3.67 | 4 | 3 | 3.6 | 4 |
| | Avg. | 2 | 3.47 | 4 | 2 | 3.57 | 4 | 3 | 3.5 | 4 |
| | Bad | 3 | 3.2 | 4 | 3 | 3.23 | 4 | 3 | 3.27 | 4 |
| Sweet-JPL | Good | 3 | 4.17 | 5 | 3 | 4 | 5 | 4 | 4.33 | 5 |
| | Avg. | 3 | 3.77 | 5 | 3 | 3.77 | 5 | 3 | 3.57 | 4 |
| | Bad | 3 | 4.13 | 5 | 3 | 3.7 | 5 | 3 | 3.9 | 5 |
| miniTambis | Good | 5 | 5.73 | 7 | 5 | 5.57 | 7 | 5 | 5.53 | 6 |
| | Avg. | 5 | 5.47 | 7 | 5 | 5.7 | 7 | 5 | 5.4 | 6 |
| | Bad | 4 | 5.67 | 7 | 5 | 5.67 | 7 | 4 | 5.8 | 7 |
| University | Good | 5 | 6.23 | 8 | 5 | 6.13 | 8 | 5 | 6.17 | 8 |
| | Avg. | 5 | 6.33 | 8 | 4 | 6.3 | 10 | 5 | 6.27 | 8 |
| | Bad | 5 | 7.13 | 10 | 5 | 6.67 | 9 | 5 | 6.67 | 8 |
| Economy | Good | 5 | 10.87 | 28 | 5 | 12.63 | 42 | 7 | 13.6 | 19 |
| | Avg. | 6 | 14.85 | 30 | 7 | 13.47 | 26 | 7 | 14.8 | 27 |
| | Bad | 7 | 16.1 | 39 | 9 | 16.42 | 36 | 9 | 17.47 | 33 |
| Transportation | Good | 5 | 10.87 | 32 | 5 | 13.53 | 26 | 5 | 14.5 | 26 |
| | Avg. | 5 | 13.57 | 27 | 5 | 14.07 | 26 | 5 | 15.4 | 22 |
| | Bad | 8 | 16.5 | 32 | 11 | 17.67 | 32 | 6 | 18.6 | 31 |

Table 9: Minimum, average and maximum number of queries required by the entropy-based and "Split-in-half" query selection methods to identify the target diagnosis in a real-world ontology. Ontologies are ordered by the number of diagnoses.

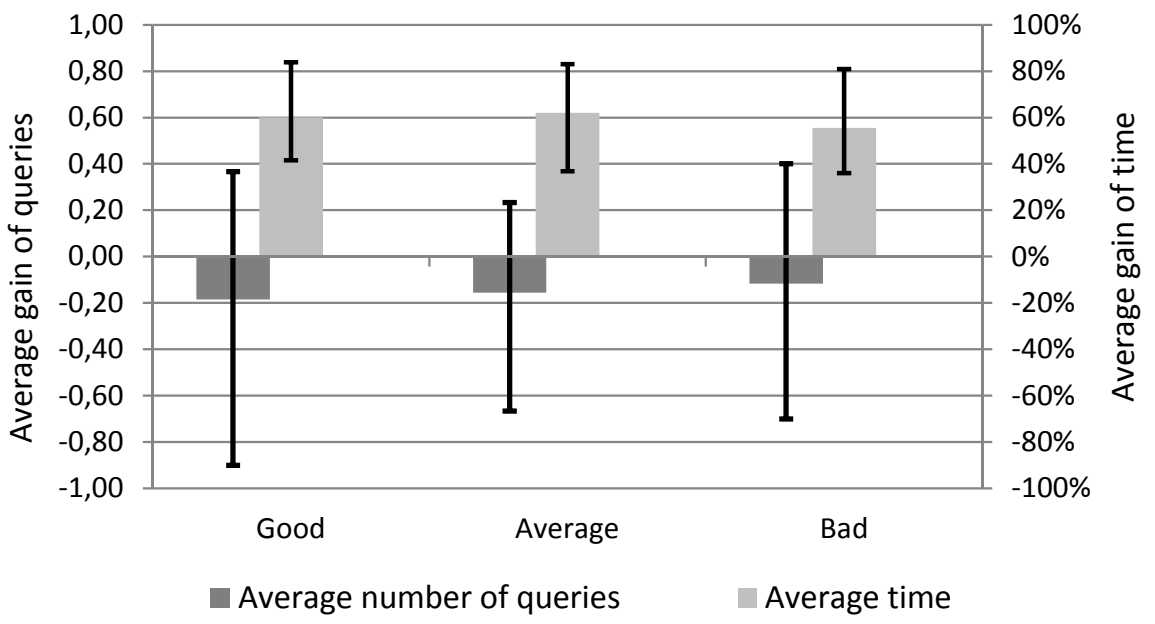

Figure 5: Average time/query gain resulting from the application of the extended CKK partitioning algorithm. The whiskers indicate the maximum and minimum possible gain of queries/time by using extended CKK.

All these measures are evaluated for each axiom in a conflict set. The scores are then combined in a rank value which is associated with the corresponding axiom. These ranks are then used by a modified HS-Tree algorithm that identifies diagnoses with a minimal rank. In this work no query generation and selection strategy is proposed if the target diagnosis cannot be determined reliably with the given a-priori knowledge. In our work additional information is acquired until the target diagnosis can be identified with confidence. In general, the work of [10] can be combined with the one presented in this paper as axiom ranks can be taken into account together with other observations for calculating the prior probabilities of the diagnoses.

The idea of selecting the next best query based on the expected entropy was exploited in the generation of decisions trees [20] and further refined for selecting measurements in the model-based diagnosis of circuits [15]. We extended these methods to query selection in the domain of ontology debugging.

## 7. Conclusions

In this paper we presented an approach to the sequential diagnosis of ontologies. We showed that the axioms generated by classification and realization can be exploited to generate queries which differentiate between diagnoses. To rank the utility of these queries we employ knowledge about typical user errors in ontology axioms. Based on the likelihood of an ontology axiom to contain an error we predict the information gain produced by a query result, enabling us to select the next best query according to a one-step-lookahead entropy-based scoring function. We outlined the implementation of a sequential debugging algorithm and compared our proposed method with a “Split-in-half” strategy. Our experiments showed a significant reduction in the number of queries required to identify the target diagnosis. In addition, our evaluation employing real-word ontologies indicates that even a rough estimate of the prior probabilities of faults with a moderate variance allow the advantageous application of the entropy-based query selection.